\documentstyle[preprint,pre,aps]{revtex}

\begin{document}

\draft
\input epsf

\title{Approximate Analytic Solution for the Spatiotemporal
Evolution of Wave Packets undergoing Arbitrary Dispersion}

\author{Borge Nodland}

\address{Department of Physics and Astronomy, and Rochester Theory
Center for Optical Science and Engineering, University of Rochester,
Rochester, New York 14627}

\date{Published in Physical Review E {\bf 55,} 3647 (1997)}

\maketitle

\begin{abstract}
We apply expansion methods to obtain an approximate expression in terms
of elementary functions for the space and time dependence of wave
packets in a dispersive medium. The specific application to pulses in a
cold plasma is considered in detail, and the explicit analytic formula
that results is provided. When certain general initial conditions are
satisfied, these expressions describe the packet evolution quite well.
We conclude by employing the method to exhibit aspects of dispersive
pulse propagation in a cold plasma, and suggest how predicted and
experimental effects may be compared to improve the theoretical
description of a medium's dispersive properties.
\end{abstract}


\section{Introduction}

A systematic study of the propagation of electromagnetic pulses in a
dispersive medium originated with Sommerfeld \cite{som1,som2} and
Brillouin \cite{bri1,bri2}. The subject has subsequently been studied
by means of a number of interesting methods and novel interpretations;
see for example references \onlinecite{and,tan,mus,she} and references
therein. Some of these methods are extensions to the classic
Sommerfeld-Brillouin theory.

The purpose of this paper is firstly to provide an analytic expression
for the behavior of dispersive wave packets by use of a straightforward
method [see Eq. (\ref{eq8})], along with a description of the
initial conditions that must be satisfied in order to render this
expression valid.  Secondly, we furnish the explicit formula that
results when the method is applied to the case of Gaussian pulses in a
cold plasma [see Eq. (\ref{eq13})]. We conclude by using the
method to discuss and exhibit features of the dispersive propagation
and interaction of multiple pulses in such a plasma.

\section{General Method}

The method we employ is based on Taylor expansion techniques and
Gaussian wave packet expansion techniques that allow required integrals
to be computed exactly.  More specifically, we expand an initial wave
packet as a superposition of Gaussian packets; then power-expand the
dispersion relation to second order around the dominant wavenumber of
each Gaussian.  This allows for an exact solution of the Fourier
integral describing the evolution of the packet.

We take as our starting point the standard linear Fourier integral
\cite{whi,jac} describing the propagation of a pulse $E(x, t)$ in a
medium of dispersion $\omega(k)$:

\begin{equation}
E(x, t) = \int_{-\infty}^\infty A(k) 
e^{i (k x - \omega(k) t)} \, dk,
\label{eq1}
\end{equation}
where $A(k)$ must be chosen to satisfy the initial boundary conditions
for $E(x, t)$ \cite{whi}. For simplicity, we assume that $\frac{d}{dt}
E(x, 0) = 0$, so that $A(k)$ is just the Fourier transform of the
initial pulse $E(x, 0) = E_0(x)$ \cite{whi,jac},

\begin{equation}
A(k) = \frac{1}{2 \pi} \int_{-\infty}^\infty 
E_0(x) e^{-i k x} \, dx.
\label{eq2}
\end{equation}
We then take the real part of Eq. (\ref{eq1}), assume $A(k)$ is
real, and denote $\text{Re} \, E(x, t)$ by $E(x, t)$ for simplicity

\begin{equation}
E(x, t) = \int_{-\infty}^\infty A(k) 
\cos\left(k x - \omega(k) t\right)  \, dk.
\label{eq3}
\end{equation}

The next step is to find a fit to $E_0(x)$ in the form of a
superposition of sinusoidal wave packets with a Gaussian envelope

\begin{equation}
E_0(x) = \sum_{i=1}^n f_i e^{-a_i x^2} \cos(b_i x),
\label{eq4}
\end{equation}
where the constants $a_i$, $b_i$, $f_i$ are real, and $a_i > 0$.  Such
a fit can normally be found for a localized initial packet $E_0(x)$.
The Fourier transform of Eq. (\ref{eq4}) can then be found exactly
as

\begin{equation}
A(k) = \frac{1}{2 \pi} 
\int_{-\infty}^\infty E_0(x) e^{-i k x} \, dx 
= \frac{1}{4 \sqrt{\pi}} \sum_{i=1}^n \frac{f_i}{\sqrt{a_i}}
\left[ e^{(\frac{-1}{4 a_i}) (k - b_i)^2} 
+ e^{(\frac{-1}{4 a_i}) (k + b_i)^2} \right].
\label{eq5}
\end{equation}
This amplitude is real, so the assumption below Eq. (\ref{eq2}) is
valid.  The dominant wave numbers in Eq. (\ref{eq5}) are $\pm
b_i$, $i=1,n$, so we expand the dispersion relation $\omega(k)$ to
second order around each point $\pm b_i$, $i=1,n$, separately:

\begin{equation}
\omega(k, \pm b_i) = s_0(\pm b_i) + 
k s_1(\pm b_i) + k^2 s_2(\pm b_i). 
\label{eq6}
\end{equation}
Equation (\ref{eq6}) constitutes $2 n$ expansions of $\omega(k)$, each
expansion being centered around one of the $2 n$ points $\pm b_i$,
$i=1,n$. The coefficients $s_j(\pm b_i)$ will in general depend on the
parameters describing the dispersion relation $\omega(k)$ of the
dispersive medium.

We can now obtain a good approximation to Eq. (\ref{eq3}) by using
Eqs. (\ref{eq5}) and (\ref{eq6}) to expand the integrand in
Eq. (\ref{eq3}) as a sum of $2 n$ terms, where each term is
centered on the major wave numbers $\pm b_i$, $i=1,n$ of the initial
packet:

\begin{eqnarray}
E(x, t) &=& \frac{1}{4 \sqrt{\pi}} 
\sum_{i=1}^n \frac{f_i}{\sqrt{a_i}}
\Biggl\{ \int_{-\infty}^\infty 
e^{(\frac{-1}{4 a_i}) (k - b_i)^2}
\cos \left[ -t s_0(b_i) + k \left(x - t s_1(b_i) \right) 
- t k^2 s_2(b_i)\right] \, dk 
\nonumber \\
&+& (b_i \rightarrow -b_i) \Biggr\} 
\label{eq7} 
\end{eqnarray}
Note that Eq. (\ref{eq7}) arises from a generalized expansion
approach to Eq. (\ref{eq3}), since it incorporates an arbitrary
number of {\it different} expansion points $\pm b_i$. The integrand in
Eq. (\ref{eq3}) is expanded to a sum of $2 n$ Gaussian terms by
substituting Eq. (\ref{eq5}) into Eq. (\ref{eq3}), and the
only modification of a term occurs via a Taylor expansion (\ref{eq6})
of the factor $\omega(k)$ around the point where the term is most
influential. The advantage of the expansion (\ref{eq7}) is 
that the resulting integrals can now be found analytically as

\begin{eqnarray}
E(x, t) &=& \frac{1}{4} \sum_{i=1}^n \frac{f_i}{\sqrt{a_i}} 
\Biggl\{ F^{\frac{1}{4}}(t, a_i, b_i) 
\nonumber \\
& & \times \exp \Biggl( F(t, a_i, b_i) (4 a_i)^{-1} 
\Biggl[\left[x - t s_1(b_i)\right] t b_i s_2(b_i) -\frac{1}{4} 
\left[x - t s_1(b_i) \right]^2
- t^2 b_i^2 s_2^2(b_i) \Biggr] \Biggr) 
\nonumber \\
& & \times \cos \Biggl( F(t, a_i, b_i) \Biggl[ (4 a_i)^{-2} 
\left\{ \left[
x - t s_1(b_i)\right] b_i - t \left[ b_i^2 s_2(b_i) + s_0(b_i) 
\right] \right\}
\nonumber \\
&+& \frac{1}{4} \left[x - t s_1(b_i)\right]^2 t s_2(b_i)
- t^3 s_2^2(b_i) s_0(b_i) \Biggr]
-\frac{1}{2} \tan^{-1}\left[4 t a_i s_2(b_i)\right] \Biggr) 
\nonumber \\
&+& (b_i \rightarrow -b_i) \Biggr\},
\label{eq8}
\end{eqnarray}
where $F(t, a_i, b_i) = \left[ (4 a_i)^{-2} + t^2
s_2^2(b_i)\right]^{-1}$. Equation (\ref{eq8}) is useful in the sense
that it is an explicit formula containing only elementary functions,
and in the sense that it is quite general. All one needs to do to apply
it to a specific situation is to compute the power series of the
dispersion relation around the main wave numbers of the initial wave
packet. The accuracy of the expression (\ref{eq8}) will of course
increase as the number of expansion points are increased. The formula
provides for expansions of $\omega(k)$ up to second order, but one
could also of course limit the order to unity by taking $s_2(\pm b_i) =
0$ in Eq. (\ref{eq8}).

We now discuss the initial conditions that must hold for Eq.
(\ref{eq8}) to be valid. From Eq. (\ref{eq5}), we see that the
amplitudes of the two terms of the integrand in the contribution

\begin{equation}
\frac{1}{4 \sqrt{\pi}} \frac{f_i}{\sqrt{a_i}}
\int_{-\infty}^\infty 
\left[ e^{(\frac{-1}{4 a_i}) (k - b_i)^2} 
+ e^{(\frac{-1}{4 a_i}) (k + b_i)^2} \right] 
\cos\left(k x - \omega(k) t\right)  \, dk
\label{eq9}
\end{equation}
to $E(x,t)$ in Eq. (\ref{eq3}) are maximal at the points $k=b_i$
and $k=-b_i$. The amplitudes decay exponentially as $k$ moves away from
the points $\pm b_i$, and are $e^{-1}$ of their maximal value when $k$
is at a distance $2 \sqrt{a_i}$ from either point. At a distance $5
\sqrt{a_i}$, the amplitudes are $e^{-6.25} \approx 1.9 \times 10^{-3}$
of their maximal value, which is so small that beyond this distance one
may consider the contribution (\ref{eq9}) to be negligible. The
$k$-intervals

\begin{equation}
I(b_i)=[b_i - 5\sqrt{a_i}, b_i + 5\sqrt{a_i}], \; 
I(-b_i)=[-b_i - 5 \sqrt{a_i},-b_i + 5 \sqrt{a_i}]
\label{eq9.3}
\end{equation}
may therefore be taken as the only two areas of the $k$-line where the
contribution (\ref{eq9}) to (\ref{eq3}) is appreciable.

In order for Eq. (\ref{eq8}) to be applicable, one must require
that a contribution (\ref{eq9}) (for a specific $i$) is appreciable
only within the two areas of the $k$-line where the two second order
expansions $\omega(k, \pm b_i)$ of $\omega(k)$ (for the above $i$) in
Eq. (\ref{eq6}) are valid. The remainders $|\omega(k) - \omega(k,
b_i)|$ and $|\omega(k) - \omega(k, -b_i)|$ will in general be small
over a certain interval containing $b_i$ and another interval
containing ($-b_i$), respectively. We denote the largest two such
intervals over which $|\omega(k) - \omega(k, b_i)|$ and $|\omega(k) -
\omega(k, -b_i)|$ are small by $I'(b_i)$ and $I'(-b_i)$ respectively.
These intervals depend on $\omega(k)$ and $\pm b_i$ [The remainder
terms depend on the third derivative of $\omega(k)$, as is well known
from Taylor expansion theory]. The requirement for Eq.
(\ref{eq8}) to be valid can accordingly be stated as

\begin{equation}
I(b_i) \subset I'(b_i), \; I(-b_i) \subset I'(-b_i), \; \; (i=1,n).
\label{eq9.5}
\end{equation}

Therefore, when the intervals $I'(b_i)$ and $I'(-b_i)$ have been
determined from $\omega(k)$ and $\pm b_i$, the domain of validity of
Eq. (\ref{eq8}) can be stated as an upper limit on the pulse shape
parameters $a_i$ in Eq. (\ref{eq4}), as noted from Eq.
(\ref{eq9.3}). Note that the $n$ criteria in Eq. (\ref{eq9.5}) are
independent, since the resultant packet $E(x,t)$ in Eq.
(\ref{eq3}) is a linear combination of $n$ independent contributions of
the form (\ref{eq9}), as seen from Eq. (\ref{eq5}).

\section{Cold Plasma Application}

We now consider the specific case of a pulse propagating in a cold
plasma. For simplicity, we assume that the initial pulse can be
approximated well by a single normalized Gaussian packet centered at
$x=0$, 

\begin{equation}
E_0(x) = e^{-a x^2} \cos(b x).
\label{eq10}
\end{equation}
In this case, Eq. (\ref{eq8}) must be applied with the parameters

\begin{equation}
n=1, \; a_i=a, \; b_i=b, \; f_i=1.
\label{eq10.5}
\end{equation}
For the cold plasma, we use the dispersion relation \cite{kra}
\begin{equation}
\omega(k) = \sqrt{c^2k^2+ \omega_p^2},
\label{eq11}
\end{equation}
where $c$ is the speed of light in vacuum, and $\omega_p$ is the plasma
frequency.

We then expand Eq. (\ref{eq11}) around $b$ and $-b$, to obtain the
coefficients in Eq. (\ref{eq6}) as

\begin{eqnarray}
s_0(b) &=& \frac{3}{2}(c^2 \omega_p^2 b^2 + 2 \omega_p^4) 
B^{\frac{3}{2}}, \;
s_1(b) = c^4 b^3 B^{\frac{3}{2}}, \;
s_2(b) = \frac{1}{2} c^2 \omega_p^2 B^{\frac{3}{2}}, 
\nonumber \\
s_0(-b) &=& s_0(b), \; s_1(-b) = -s_1(b), \; s_2(-b) = s_2(b),
\label{eq12}
\end{eqnarray}
where $B=(b^2 c^2 + \omega_p^2)^{-1}$. These coefficients are
general, since they are functions of the initial pulse parameter
$b$.

Substitution of Eqs. (\ref{eq10.5}) and (\ref{eq12}) in Eq.
(\ref{eq8}) yields

\begin{eqnarray}
E(x, t) &=& \frac{1}{4} a^{-\frac{1}{2}} T^{\frac{1}{4}}
\exp \{ (x - t b^3 c^4 B^{\frac{3}{2}}) 
t b c^2 \omega_p^2 (8 a)^{-1} B^{\frac{3}{2}}
-(x - t b^3 c^4 B^{\frac{3}{2}})^2 (16 a)^{-1} T 
\nonumber \\
&-&t^2 b^2 c^4 \omega_p^4 (16 a)^{-1} B^3 \} 
\nonumber \\
&\times& \cos \{ T [ 
\frac{1}{16} (x - t b^3 c^4 B^{\frac{3}{2}}) a^{-2} b
+ \frac{1}{8} (x - t b^3 c^4 B^{\frac{3}{2}})^2 t c^2 
\omega_p^2 B^{\frac{3}{2}}
\nonumber \\
&-& \frac{1}{16} t a^{-2} B^{\frac{3}{2}}
(2 b^2 c^2 \omega_p^2 + \omega_p^4) 
- \frac{1}{8} t^3 \omega_p^4 c^4 B^{\frac{9}{2}}
(3 b^2 c^2 \omega_p^2 + 2 \omega_p^4) ]
\nonumber \\
&-& \frac{1}{2} \tan^{-1}(2 t a c^2 \omega_p^2 
B^{\frac{3}{2}}) \} + (b \rightarrow -b),
\label{eq13}
\end{eqnarray}
where $T=(\frac{1}{16} a^{-2} + \frac{1}{4} t^2 c^4 \omega_p^4
B^3)^{-1}$ and $B=(b^2 c^2 + \omega_p^2)^{-1}$. 
Equation (\ref{eq13}) is a general equation, describing the dispersive
behavior of pulses of the form (\ref{eq10}) that propagate in a medium
with dispersion of the form (\ref{eq11}).

The intervals $I(\pm b_i)$ and the validity criteria for Eq.
(\ref{eq13}) are found by substitution of Eq. (\ref{eq10.5})
in Eqs. (\ref{eq9.3}) and (\ref{eq9.5}). Since the dispersion
relation $\omega(k)$ in Eq. (\ref{eq11}) is symmetric about $k=0$,
the validity criteria in Eq. (\ref{eq9.5}) simplify.  A second
order expansion $\omega(k, b)$ of the square root $\omega(k)=\sqrt{c^2
k^2 + \omega_p^2}$ about a positive wave number $b$ approximates
$\omega(k)$ quite well when $k>0$, but fails when $k<0$. One must
therefore require the interval $I(b)$ in Eq. (\ref{eq9.3}) to lie
to the right of the origin of the $k$-line.  According to Eq.
(\ref{eq9.3}), the left endpoint of $I(b)$ is to the right of the
origin when $b>5 \sqrt{a}$. $I(b)$ then lies within $[0, b+5
\sqrt{a}]$, which is within the interval $I'(b)$ of validity of the
expansion $\omega(k, b)$ [given by Eqs. (\ref{eq6}) and
(\ref{eq12})] of $\omega(k)$ [given by Eq. (\ref{eq11})].
A similar result holds for the expansion $\omega(k, -b)$ of
$\sqrt{c^2 k^2 + \omega_p^2}$ about a negative wave number $-b$.
Hence, we have the useful initial condition that

\begin{equation}
\frac{b}{\sqrt{a}} > 5
\label{eq14}
\end{equation}
for Eq. (\ref{eq13}) to be valid. In other words, the
approximation (\ref{eq13}) is accurate when the frequency $b$ of the
initial pulse $E_0(x)$ in Eq. (\ref{eq10}) is high, or when
$E_0(x)$ is broad (i.e. when the pulse shape parameter $a$ is small).
However, we see from Eq. (\ref{eq14}) that short pulses are also
well approximated by Eq. (\ref{eq13}) when they are of high
frequency, as well as low-frequency pulses of long duration. 

In the specific illustrations and discussions below, the pulse
parameters $a$ and $b$ were of course always chosen so that Eq.
(\ref{eq14}) was satisfied, with many choices such that
$\frac{b}{\sqrt{a}} >> 5$. As a double check, we also compared values
for $E(x, t)$ given by Eq. (\ref{eq13}) with the values for $E(x,
t)$ obtained by numerically integrating Eq. (\ref{eq3}) [using
Eqs. (\ref{eq5}), (\ref{eq10}), (\ref{eq10.5}) and (\ref{eq11})]
for a systematic variation of values for $x,t,a,b,c$ and $\omega_p$.
For values of $a$ and $b$ obeying Eq. (\ref{eq14}), the agreement
was excellent, as expected. The discrepancy between the exact and
approximate values for $E(x, t)$ was typically found to be less than
$10^{-4}$.

Equation (\ref{eq13}) represents two pulses traveling in opposite
directions, as we expect \cite{jac} from the initial condition
$\frac{d}{dt} E(x, 0) = 0$, stated above Eq. (\ref{eq2}).  For
simplicity, in the remaining part of this paper we focus only on the
right-traveling pulse term in Eq. (\ref{eq13}) (which is the first
term in Eq. (\ref{eq13}) when $b$ is positive).

We now outline some of the features of the dispersive behavior of a
single pulse in a cold plasma, which can be observed from plots of
Eq. (\ref{eq13}). It is readily observed that the envelope of a
short initial pulse broadens more rapidly as time elapses compared to the
broadening of the envelope of a long initial pulse, the initial pulse
frequencies being equal. This is of course well known, and can be
explained analytically \cite{jac}.

It is also seen that in general, the number of oscillations within the
pulse envelope increases with time, and more interestingly, the effect
of dispersion on the pulse is to shift high-frequency components of the
pulse toward the spatial front of the pulse and low-frequency
components of the pulse toward the spatial rear of the pulse (see the
pulse $E1$ in Fig. \ref{fig4}). This effect becomes more prominent as
the initial pulse is shortened.

Additionally, the spatial packet velocity of a pulse increases as the
pulse frequency is increased (with spatial packet velocity we mean
$x_{max}/t$, where $x_{max}$ is the distance traveled by the spatial
maximum of the pulse envelope during a given time period $t$.  Temporal
pulse velocity is similarly defined as $x/t_{max}$, where $t_{max}$ is
the time that elapses before the temporal maximum of the pulse envelope
appears at a given distance $x$ \cite{and}). However, as expected, in
no cases is the center of a packet beyond the point $x=c t$ at
time $t$ if it is at the origin $x=0$ at time $t=0$. Here $c$ is the
velocity of light in vacuum, appearing in Eq. (\ref{eq13}) from
the dispersion relation (\ref{eq11}).

Moreover, the envelope shape is better maintained for a high frequency
initial pulse than for a low frequency initial pulse, the initial pulse
durations being equal. Therefore, increasing the frequency of a short
pulse will improve the preservation of its shape as it travels through
a dispersive medium, as well as increase its packet speed. The
propagation of short, high-frequency pulses is a topic of strong
current interest \cite{dec}, \cite{fei}.

We next examine the the dispersive evolution of more than one
(right-traveling) pulse through a cold plasma. One can readily do this
by repetitive use of Eq. (\ref{eq13}). For example, the
propagation of two initial pulses $E_{01}(x) = e^{-a_1 x^2} \cos(b_1
x)$ and $E_{02}(x) = e^{-a_2 x^2} \cos(b_2 x)$ separated by an initial
time delay $d$ is described by

\begin{equation}
E(x,t)=E1(x,t) + E2(x,t),
\label{eq15}
\end{equation}
where $E1(x,t)$ and $E2(x,t)$ are obtained by substituting ($t
\rightarrow t+d$, $a \rightarrow a_1$, $b \rightarrow b_1$) and ($a
\rightarrow a_2$, $b \rightarrow b_2$) respectively, into Eq.
(\ref{eq13}). The first pulse $E1$ is here $d$ time units ahead of the
second pulse $E2$ when $E2$ passes the origin $x=0$ at the time $t=0$
(A description of the general dispersive behavior of $m$ initial pulses
$E_{0j}$ of the form (\ref{eq4}) separated by initial time delays $d_j$
can analogously be obtained from Eq. (\ref{eq8}) by the set of
substitutions $t \rightarrow t+d_j$, $a_i \rightarrow a_{ij}$, $b_i
\rightarrow b_{ij}$ for $i=1,n$ and $j=1,m$ ).

Figures \ref{fig1} through \ref{fig4} depict some features of
dispersive double pulse propagation, as described by Eqs.
(\ref{eq13}) and (\ref{eq15}). These figures also illustrate some of
the properties of single pulse propagation mentioned above (note that
the packets in the top plots of Figs. \ref{fig1} and \ref{fig3} at
the initial time $t=0$ are identical to those given by Eq.
(\ref{eq10}), except that their amplitudes have half the magnitude of
those in Eq. (\ref{eq10}). This is because half of the energy of
the initial packets is carried by the left-traveling parts of $E(x,
t)$, which are omitted from all plots).

Figure \ref{fig1} shows how two initially distinct packets of the same
frequency and duration gradually overlap and interact with each other
due to the dispersive stretching of the two packets as time elapses.
Figure \ref{fig2} shows the results of the interaction after a very
long time. One can clearly see the emergence of constructive and
destructive interference effects among the different spatial regions of
the coalescing pulses.

In Fig. \ref{fig3}, a high frequency, short pulse is dispatched after
the dispatch of a low frequency, long pulse. We see that the high
frequency pulse eventually overtakes the low frequency pulse, in accord
with the single-pulse feature that the packet velocity is greater for a
high-frequency initial pulse than a low-frequency initial pulse. The
middle and bottom plots show how the high frequency pulse interferes
with the various spatial parts of the low frequency pulse. The bottom
plot depicts an instant of constructive interference between the two
pulses, where the total amplitude is larger than the separate
amplitudes of either pulse. For longer times than those shown in the
figure, the high frequency pulse eventually passes and leaves the other
pulse.

As a final example of a usage of Eq. (\ref{eq13}), one may employ
this Eq. to determine parameters $a_j$, $b_j$, and initial time
delays $d_j$, $(j=i,m)$ of $m$ pulses such that they will all have the
same width and completely overlap at a specified time $t$ while
propagating in a plasma of dispersion characterized by $c$, $\omega_p$
and Eq. (\ref{eq11}). Figure \ref{fig4} shows a two-pulse version
of this, with $t=1000$. We see that strong destructive and constructive
interference effects occur in this case. A comparison of such predicted
effects with those found experimentally in a specific medium could be
used to determine the medium's dispersion relation $\omega(k)$.

\section{Summary and Conclusions}

We have provided an analytic expression describing the propagation of
dispersive wave packets [Eq. (\ref{eq8})] provided the packets
satisfy the applicability criteria (\ref{eq9.5}). The expression is
obtained by a clear-cut method, and can be used to study properties of
the propagation process (for example pulse velocity and multiple pulse
interference effects due to dispersion).

In the specific case of packets propagating in a cold plasma, we used
Eq. (\ref{eq8}) with two terms, and provided the explicit,
analytic expression that results in this case [Eqs. (\ref{eq13})
and (\ref{eq15})]. For initial pulses satisfying the condition
(\ref{eq14}), Eq. (\ref{eq13}) gives an accurate description of
single or multiple pulse propagation in a dispersive cold plasma. 

As a final note, it is conceivable that the presence or non-presence of
effects predicted by Eqs. (\ref{eq8}) and (\ref{eq13}) could be
used to verify, improve the equations for, or determine a substance's
dispersive properties.

\acknowledgements

This research was supported by NSF Grant No. PHY94-15583. The numerical
part of the work was done at the Laboratory for Laser Energetics.

\begin{figure}
\caption{Temporal evolution of two pulses $E1$ (right) and $E2$ (left)
in a cold plasma, obtained from Eqs. (\protect\ref{eq13}) and
(\protect\ref{eq15}). The initial time delay between the two packets is
$d=1.5$. Both pulses are identical, with initial-value parameters
$a_1=a_2=10$ and $b_1=b_2=20\frac{\omega_p}{c}$. The plasma is
characterized by $c=1$ and $\omega_p=1$. The spatial distribution of
the total field $E(x,t)=E1(x,t)+E2(x,t)$ is shown at times $t=0$ (top),
$t=500$ (middle) and $t=1000$ (bottom).}
\label{fig1}
\end{figure}

\begin{figure}
\caption{Same as in Fig. \protect\ref{fig1}, but for $t=5000$ (top)
and $t=100000$ (bottom).}
\label{fig2}
\end{figure}

\begin{figure}
\caption{Temporal evolution of two pulses $E1$ (right) and $E2$ (left)
in a plasma, where $E1$ is characterized by $a_1=5$,
$b_1=20\frac{\omega_p}{c}$, $E2$ by $a_2=100$,
$b_2=60\frac{\omega_p}{c}$, and the plasma by $c=1$, $\omega_p=1$ and
Eq. (\protect\ref{eq11}). $E1$ is initially $d=1.1$ time units
ahead of $E2$. The spatial distribution of the total field
$E(x,t)=E1(x,t)+E2(x,t)$ is shown at times $t=0$ (top), $t=500$
(middle) and $t=1000$ (bottom).}
\label{fig3}
\end{figure}

\begin{figure}
\caption{Spatial dependence of a packet $E1(x,t)$ (top) and a packet
$E2(x,t)$ (middle) after they have both traversed a cold plasma
(characterized by $c=1$, $\omega_p=1$ and Eq.
(\protect\ref{eq11})) for a period of $t=1000$ time units. Initially,
$E1$ was $d=1.1$ time units ahead of $E2$. $E1$ is characterized by
$a_1=10$, $b_1=20\frac{\omega_p}{c}$ and $E2$ by $a_2=2$,
$b_2=40\frac{\omega_p}{c}$. The bottom plot shows the total field
$E(x,1000)=E1(x,1000)+E2(x,1000)$.}
\label{fig4}
\end{figure}
 
\newpage
\centerline{
\epsfbox[50 580 350 710]{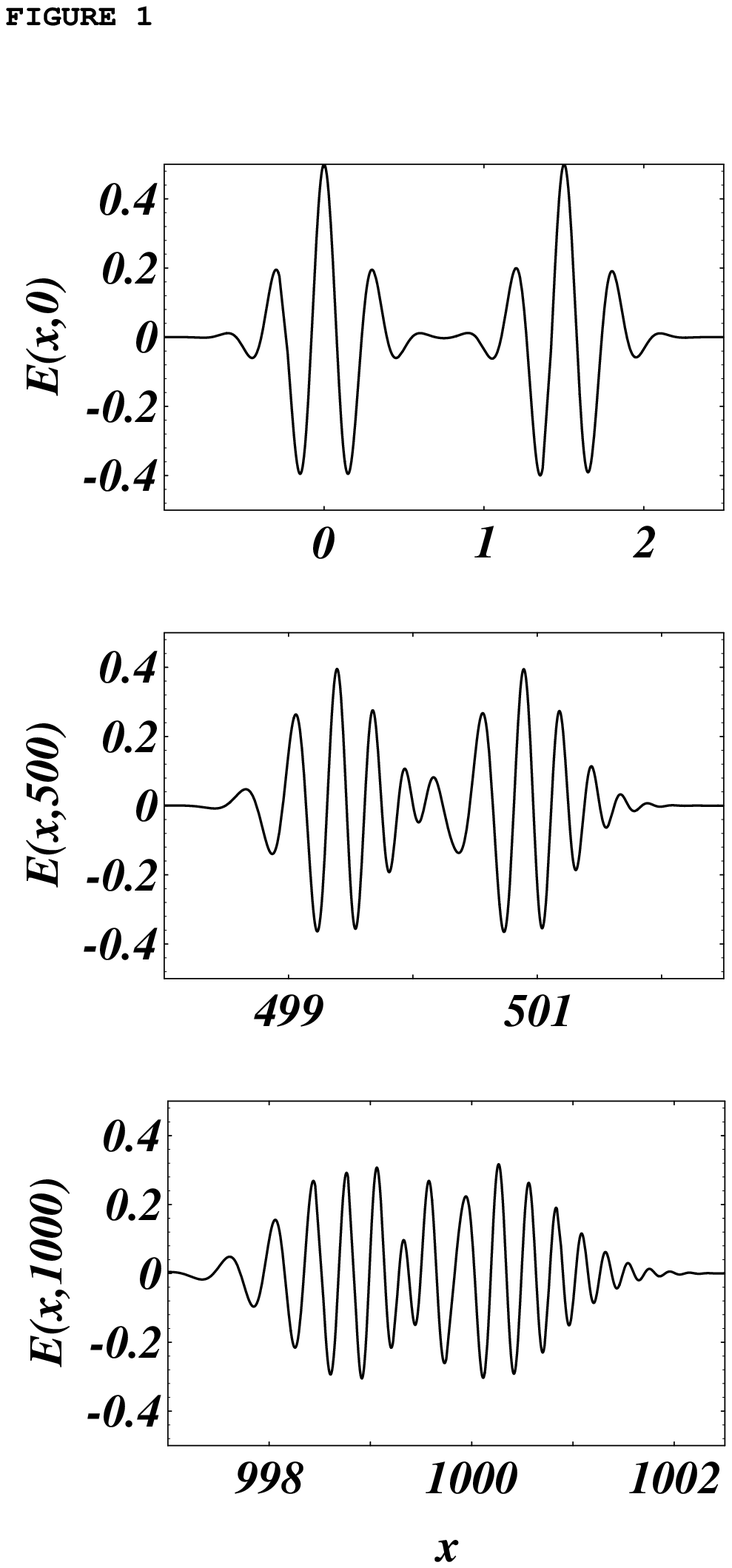} }

\newpage
\centerline{
\epsfbox[50 580 350 710]{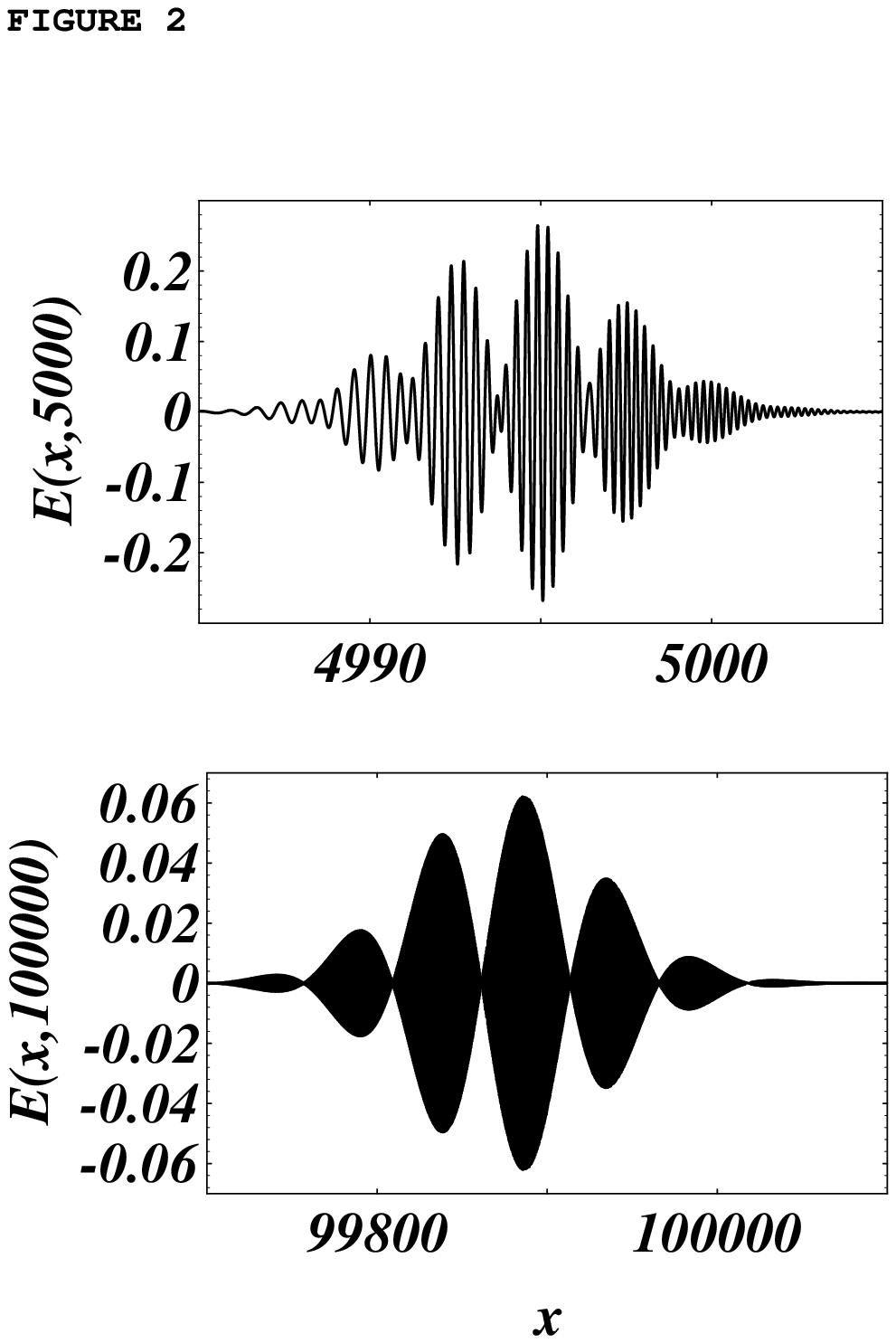} }

\newpage
\centerline{
\epsfbox[50 580 350 710]{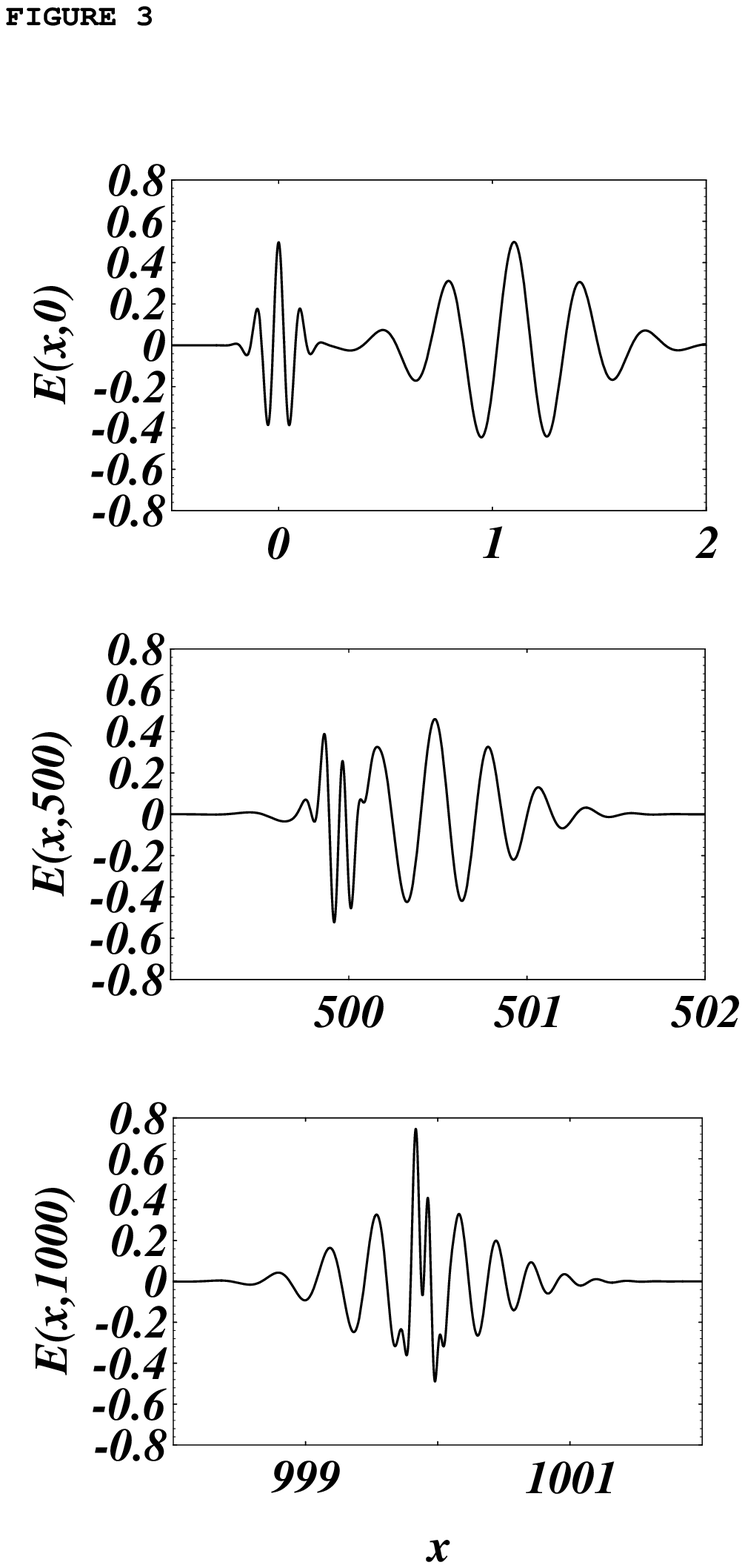} }

\newpage
\centerline{
\epsfbox[50 580 350 710]{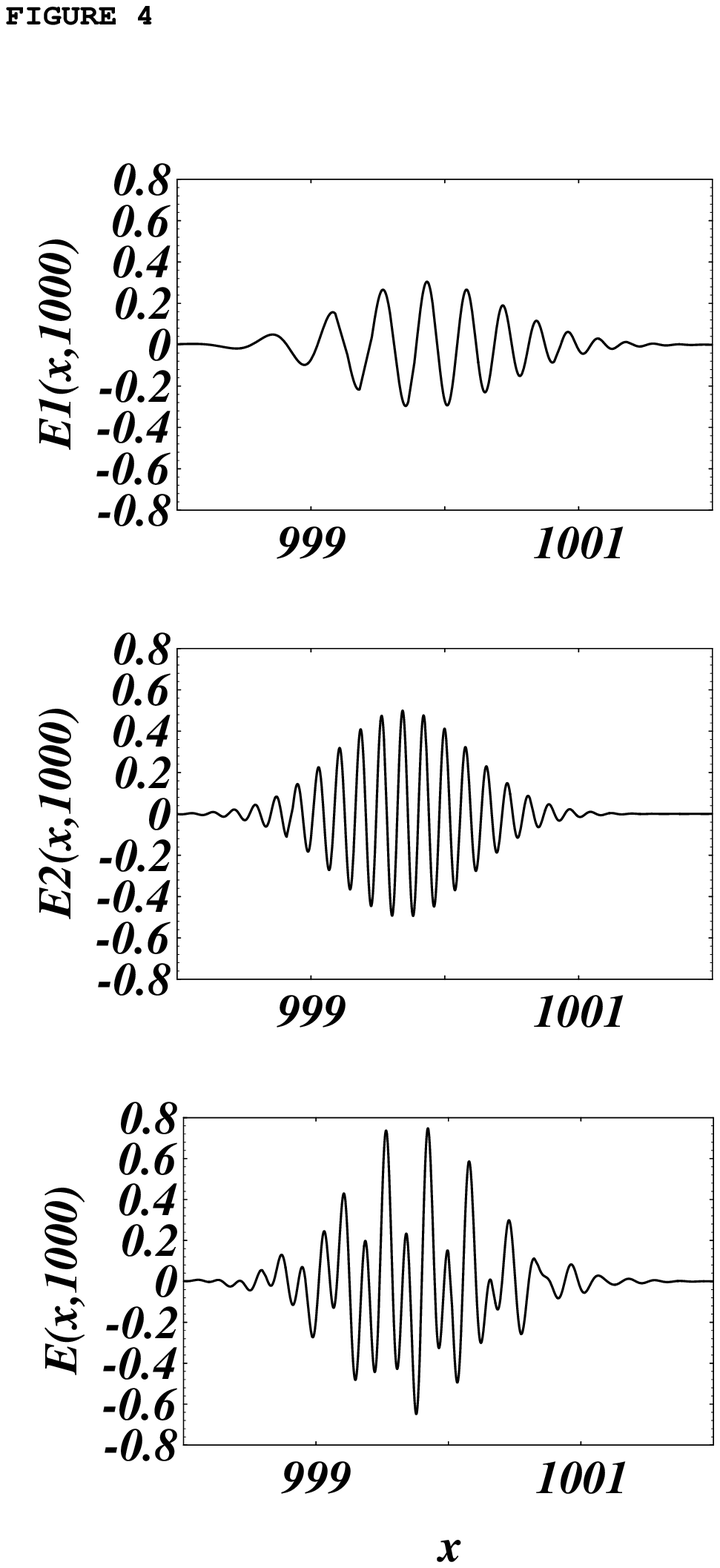} }


\begin{references}

\bibitem{som1} A. Sommerfeld, Ann. Phys. {\bf 44,} 177 (1914).

\bibitem{som2} A. Sommerfeld, {\it Optics} (Academic, New York, 1954).

\bibitem{bri1} L. Brillouin, Ann. Phys. {\bf 44,} 203 (1914).

\bibitem{bri2} L. Brillouin, {\it Wave Propagation and Group Velocity}
(Academic, New York, 1960).

\bibitem{and} D. Anderson, J. Askne and M. Lisak, Phys. Rev. A {\bf
12,} 1546 (1975).

\bibitem{tan} M. Tanaka, M. Fujiwara and H. Ikegami, Phys. Rev. A {\bf
34,} 4851 (1986).

\bibitem{mus} L. Muschietti and C. T. Dunn, Phys. Fluids B {\bf 5,}
1383 (1993).

\bibitem{she} G. C. Sherman and K. E. Oughstun, J. Opt. Soc. Am. B {\bf
12,} 229 (1995).

\bibitem{whi} G. B. Whitham, {\it Linear and Nonlinear Waves} (Wiley,
New York, 1974), Chap. 11.

\bibitem{jac} J. D. Jackson, {\it Classical Electrodynamics}, 2nd ed.,
(Wiley, New York, 1975), Chap. 7.

\bibitem{kra} N. A. Krall and A. W. Trivelpiece, {\it Principles of
Plasma Physics}, (San Francisco Press, San Francisco, 1986), p. 149.

\bibitem{dec} C. D. Decker {\it et al.}, Phys. Plasmas, {\bf 3,} 2047
(1996).

\bibitem{fei} M. D. Feit, J. C. Garrison and A. M. Rubenchik, Phys.
Rev. E, {\bf 53,} 1068 (1996).

\end{references}
\end{document}